\documentclass[aps,pre,twocolumn,groupedaddress]{revtex4}
\usepackage{amsmath}
\usepackage{color}
\usepackage{subfigure}
\usepackage[outercaption,wide]{sidecap}
\usepackage[rflt]{floatflt}
\usepackage[dvips]{graphicx}
\newcommand \el{{\emph{l}}}

\newcommand \bx{{\bar{x}}}
\newcommand \bL{{\bar{L}}}

\newcommand \bX{{\bar{X}}}
\newcommand \bsigma{{\bar{\sigma}}}
\newcommand \bzeta{{\bar{\zeta}}}

\begin{document}
\title{Period fissioning and other instabilities of stressed elastic membranes}
\author{Benny Davidovitch}
\affiliation{Physics Department, University of Massachusetts,
Amherst MA 01003}
\date{\today}
\begin{abstract}
We study the shapes of elastic membranes under the simultaneous
exertion of tensile and compressive forces when the translational
symmetry along the tension direction is broken. We predict a
multitude of novel morphological phases in various regimes of a
2-dimensional parameter space $(\epsilon,\nu)$ that defines the
relevant mechanical and geometrical conditions. Theses parameters
are, respectively, the ratio between compression and tension, and
the wavelength contrast along the tension direction. In particular,
our theory associates the repetitive increase of pattern
periodicity, recently observed on wrinkled membranes floating on
liquid and subject to capillary forces, to the morphology in the
regime ($\epsilon \!\ll \!1,\nu \! \gg \! 1$) where tension is
dominant and the wavelength contrast is large.
\end{abstract}
\maketitle
Thin membranes, such as paper sheets, tend to buckle when
compressive forces 
are exerted on their boundaries. The origin of this familiar
phenomenon, known as Euler instability, is the large contrast
between the energetic costs of straining and strain-free bending of
thin elastic bodies. This feature is reflected in the different
dependencies of the bending modulus $B \!\!\!\sim\!\! E t^3$ and
stretching modulus $Y\! \!\!\sim \!\! E t$ on the Young modulus $E$
and thickness $t$ of the membrane \cite{LLelasticity}. Wrinkling
patterns, which often appear on supported membranes such as human
skin or milk crusts, are characterized by a buckling scale $\el_{0}$
that can be much smaller than the membrane width $W$ in
the compression direction. Here, 
the distortion of the attached substrate gives rise to an energetic
cost that is proportional to the amplitude $\zeta_0$ of the bent
shape, and the formation of a 1-dimensional ($1d$) periodic
wrinkling pattern is induced by balancing restoring forces
associated with this distortion and with bending resistivity of the
membrane \cite{Cerda03}. Near buckling (wrinkling) threshold, the
ratio $\tilde{\Delta} \!\!=\!\! \Delta /W$ between the displacement
$\Delta$ and width $W$ is small (see Fig. 1) , and the membrane
attains a sinusoidal shape whose wavelength and amplitude are:
\begin{equation}
\text{(a)} \ \el_{0} = \sqrt{\frac{B}{|\sigma_{yy}|}} \  \ ;  \ \
\text{(b)} \ \frac{\zeta_0}{\el_{0}} =
{\sqrt{{\tilde{\Delta}}}}/{\pi} \label{generallaw}
\end{equation}
\begin{figure}
\vspace{-5mm}
\includegraphics[height=1.7in,clip=]{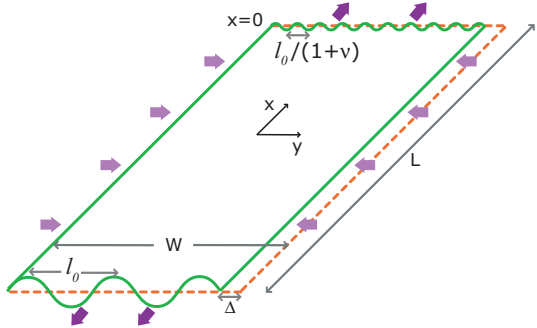}
\caption{Membrane geometry studied in this Letter.} \vspace{-5mm}
\end{figure}
where $\sigma_{yy}\!\!<\!\!0$ is the compressive stress, and the
{\emph{inextensibility}} criterion (\ref{generallaw}b) reflects the
conservation of the contour length of material lines along the
compression direction $\hat{y}$. This description corresponds to
situations in which the exerted forces are purely compressive and
enable the formation of a 1d shape. This ideal picture can be
qualitatively modified when: {\emph{(1)}} the translational symmetry
along the orthogonal direction ($\hat{x}$) is broken, \emph{or: (2)}
when tension $\sigma_{xx}\!\!=\!\!T$ is exerted on the membrane. The
dramatic effect of symmetry breaking can be demonstrated by
considering a bent membrane, whose amplitude $\zeta(x,y)$ is forced
to assume a sinusoidal shape of wavelength $\el_e \!=
\!\el_0/(1+\nu)$ at the uncompressed edge ($x\!\!=\!\!0$), where
$\nu$ is the {\emph{wavelength contrast}}. As the inextensibility
criterion (\ref{generallaw}b) shows, the limit $\nu \!\! \to
\!\!\infty$ corresponds to a vanishing amplitude $\zeta_e \to 0$ at
the edge, a geometry similar to the "curtain problem" studied by
Pomeau and Rica \cite{Pomeau}. As was shown in \cite{Pomeau}, the
transition from the vanishing amplitude at $x\!\!=\!\!0$ to the
finite value $\zeta_0$, Eq.~(\ref{generallaw}b), is characterized by
a cascade of sharp folds whose lengths may become infinitesimally
small as $x\!\to\! 0$. The important role of tension was recently
noticed by Cerda and Mahadevan \cite{Cerda03}, who showed that
exerting tension $\sigma_{xx}\!\!=\!\!T$ on a membrane with length
$L$ and free edges at $y\!\!=\!\!0,W$ implies a small compression
$|\sigma_{yy}|\!\sim\!(t/L) T$ that may induce a periodic wrinkling
pattern in the ${\hat{y}}$ direction. A useful number for
characterizing tensile effects on wrinkling phenomena is the
{\emph{stress ratio}} $\epsilon = |\sigma_{yy}|/\sigma_{xx}$.

In this Letter we classify the morphologies of elastic membranes for
general values of the pair of parameters ($\epsilon,\nu$) near
buckling (wrinkling) threshold ($\tilde{\Delta} \! \ll \! 1$). Our
theory reveals a surprisingly rich phase space, and in particular
predicts the existence of novel types of {\emph{smooth cascades}},
qualitatively different from the hierarchy of sharp folds found in
\cite{Pomeau}. Our interest in this problem was triggered by the
repetitive increase of wrinkles periodicity in discrete steps, that
was recently observed near the uncompressed edges
($x\!\!=\!\!-\!L,0$) of ultrathin membranes floating on liquid
surface and compressed along $\hat{y}$ with $\tilde{\Delta} \! \ll
\!1$ \cite{Menon08}. As was realized in \cite{Menon08}, these
membranes are under large tension $\sigma_{xx}$, and the amplitude
$\zeta_e$ at their uncompressed edges is suppressed due to the high
energetic cost of the liquid-vapor menisci (at $x\!>\!0,x\!<\!-L$),
induced by the bent membrane. The experimental conditions in
\cite{Menon08} thus correspond to the asymptotic regime $(\epsilon
\! \ll \! 1, \nu \! \gg 1)$, which is the main focus of this Letter.

Our geometry is depicted in Fig. 1. It is similar to the one studied
in \cite{Pomeau} with three crucial differences: {\bf (i)} The
displacement $\Delta$ in the $\hat{y}$ direction is assumed constant
(rather than constant compression $\sigma_{yy}\!\!=\!\!P$). {\bf
(ii)} The tension $\sigma_{xx} \!\!=\!\! T$ in $\hat{x}$ is assumed
constant (rather than constant length $L$), and {\bf (iii)} The
wavelength contrast $\nu$ can be any positive number (rather than
$\nu \!\!\to\!\! \infty$). In order to describe the wrinkling
patterns of \cite{Menon08}, we assume the membrane is floating on
liquid, and include the energetic cost of the lifted liquid mass of
density $\rho$ induced by the bent membrane. Our formalism and
results are generalizable, however, to wrinkling problems in which
gravity is replaced by another restoring force (e.g. due to an
attached elastic substrate) or if bending is the only energetic cost
(i.e. buckling). To quadratic order in $\tilde{\Delta}$, the
{{\emph{areal}}} energy density of a shape $\zeta(x,y)$ is
\cite{Menon08}:
\begin{equation}
u = \frac{1}{2}\Big(\! B (\nabla^2 \! \zeta)^2 \!+\! \rho g \zeta^2
\!+\! \sigma(x)[(\frac{\partial \zeta}{\partial y})^2  \!-\! 2
{\tilde{\Delta}}] \!+\! T(\frac{\partial \zeta}{\partial x})^2 \!
\Big) \label{energydensitybulk}
\end{equation}
where $\sigma(x)$, the compressive stress $\sigma_{yy}$ that must be
exerted at $y\!\!=\!\!0,W$ in order to impose a constant
$\tilde{\Delta}$, appears as a Lagrange multiplier that forces
{\emph{inextensibility}} of contour lines parallel to the $\hat{y}$
direction. Let us briefly review the case of a free uncompressed
edge $y=0$ \cite{Chicago08,Menon08}. The energy is then minimized by
the 1d periodic pattern:
\begin{equation}
\text{(a)} \ \zeta(x,y) = \zeta \sin(q y) \ \ ; \ \ \text{(b)} \
\zeta= \frac{2}{q} \sqrt{{\tilde{\Delta}}} \label{periodicshape}
\end{equation}
with $q\!\!=\!\!q_{0} \!\equiv \!(\rho g/B)^{1/4}$ and $\sigma_{yy}
\!\!=\!\! \sigma_{0} \!\!=\!\! -2\sqrt{B\rho g}$. The parameter
$\epsilon$ is hence defined as the ratio $\epsilon
\equiv\sigma_0/T$. The energy of the shape~(\ref{periodicshape}) is
the work $\sigma_0 \!\Delta L$, whereas the {\emph{in-plane}}
compression energy is $\frac{1}{2}Y {\tilde{\Delta}}^2 \!W \!L$. One
thus obtains the threshold value ${\tilde{\Delta}}_{\!min\!}
\!\!=\!\! 2 \sigma_{0} / Y$, below which the membrane does not bend.
Consider now the case in which the edge $x\!\!=\!\!0$ is forced to
take an unstrained periodic shape (\ref{periodicshape}b) with a
wavenumber $q_{e}\!\!=\!\!(1\!\!+\!\!\nu) q_{0}$. We assume the
length $L$ is sufficiently large such that away from the forced edge
at $x\!\!=\!\!0$ the membrane fully recovers its energetically
favorable form (\ref{periodicshape}a) with $q\!\!=\!\!q_{0}$. A
natural guess for the shape is then the superposition:
\begin{eqnarray}
\zeta (x,y) = \zeta_{0} (x) \sin(q_{0} y) + \zeta_{1}(x) \sin(q_{1}
y) \ ,
\label{2modessolution} \\
\text{where:}  \  \ q_{0}^2 \zeta_{0}^2 + q_{1}^2 \zeta_1^2 = 4
\tilde{\Delta} \
, \label{inextensibility2modes}  \\
\text{and:}  \ \ \text{(a)} \ \lim_{x \to X_0} \zeta_1(x) = 0 \  ; \
\text{(b)} \ \lim_{x \to X_1} \zeta_{0}(x) = 0 \ , \label{BC2modes}
\end{eqnarray}
where (\ref{inextensibility2modes}) is the inextensibility condition
(recall that $\tilde{\Delta} \! \ll \! 1$), and
$X_0,X_1\!\!=\!\!-\!\!L,0$, respectively, and similarly
$q_1\!\!=\!\!q_e$ are introduced to simplify the forthcoming
analysis. Notice that the superposition (\ref{2modessolution})
constitutes the {\emph most symmetric} shape possible under the
boundary conditions (BC), Eq.~(\ref{BC2modes}). Obviously, such a
smooth shape is markedly different from the irregular one described
in \cite{Pomeau} in the tensionless case ($T\!\!=\!\!0$). This
difference stems from the Gaussian curvature imposed in the membrane
by a wavelength contrast $\nu\!\!>\!\!0$, and the associated
{\emph{anharmonic}} energy density $u_G$ whose minimization (for
$T\!\!=\!\!0$) gives rise to stress focusing at ridges and vertices
which relieves the strain at all other areas of the distorted
membrane \cite{Witten07}. This principle underlies the emergence of
sharp folds in the tensionless limit $\epsilon \!\!\to\!\! 0$
\cite{Pomeau}. Assuming for the smooth shape (\ref{2modessolution})
$\partial_x \zeta_i \!\!\sim\!\! \zeta_i/\el$, with some typical
length $\el$, one obtains $u_G \!\!\sim \!\!Y {\el}^{\!-2\!}
\sum_i\!\zeta_i^4\! q_i^2$ \cite{Witten07,preparation}. For $T
\!=\!\delta \cdot Y \!\neq\! 0$ the situation is different since
even regions free from Gaussian curvature are penalized by the term
$u_T \!\!=\!\! T (\partial_x \zeta)^2 \!\!\sim\!\! T {\el}^{\!-2\!}
\sum_i\!\zeta_i^2 $. A transition from an irregular shape (at
$\epsilon \!\!\to\!\!0$) to a smooth one (\ref{2modessolution}) is
thus expected if $u_T\!>\!u_G$. With
Eq.~(\ref{inextensibility2modes}) this inequality implies:
${\tilde{\Delta}} \! \lesssim \!\delta$. Recalling the threshold
condition ${\tilde{\Delta}} \! \gtrsim \!\sigma_{0} \!/ Y$, we
obtain a necessary condition for the existence of a smooth
shape~(\ref{2modessolution}): $\delta \!\cdot\! \epsilon \! \lesssim
\! {\tilde{\Delta}} \! \lesssim  \! \delta \Rightarrow \epsilon\!
\lesssim \!1$. We thus conclude that a transition between an
irregular, sharply folded shape \cite{Pomeau} and a smooth
superposition (\ref{2modessolution}) occurs at $\epsilon^{SI} \!\!
\sim \!\! O(1)$. A more careful analysis of Euler-Lagrange (EL)
equations for the shape~(\ref{2modessolution}) reveals that
$\epsilon^{SI}$ increases with the wavelength contrast $\nu$, and
moreover $\epsilon^{SI}(\nu \!\to\!0) \geq 0.5$ \cite{preparation}.
Since for $\epsilon,\nu \!\! \to \!\!\infty$ the shape is
characterized by a diverging number of generations of sharp folds
\cite{Pomeau}, we conjecture: \vspace{2mm}

\hspace{-2mm} {\emph{{{\bf{I.}} There exists a ``branching" series
$\{b_{n}(\epsilon)\}$, such that for $(\epsilon,\nu)$ with
$\epsilon\!\!
> \!\!\epsilon^{\!SI\!}(\!\nu\!) \! \ , \ b_{\!n\!}(\epsilon\!)
\!\!<\!\! \nu \!\!<\! \!b_{\!n\!+\!1\!}(\epsilon\!)$, the
morphological phase is $I_n$, characterized by $n$ generations of
sharp folds.}}

\vspace{2mm}

Let us focus now on the asymptotic regime $(\nu , \epsilon
\!\!\ll\!\! \epsilon^{SI}(\nu))$, where a smooth shape described by
Eq.~(\ref{2modessolution}) is expected. We transform now
to the dimensionless set:
\begin{equation}
\!\bar{x} \!\!=\!\! \frac{x}{\el_T} \!;  \bar{y} \!\!=\!\!
\frac{y}{\el_T} ; \bar{\zeta} \!\!=\!\! \frac{\zeta}{2 \el_T
{\tilde{\Delta}}^{1/2}} \!; \bar{u} \!\!=\! \!\frac{u}{4
\tilde{\Delta} T \!\epsilon^{2}} \!; \bsigma \!\!=\!
\!\frac{\sigma}{T} \!; a_j \!\!=\! \!\sqrt{\epsilon}
q_{\!j}{\el_T}, \label{nondimensional}
\end{equation}
where $\el_T \!\!=\!\! \sqrt{T\!/\!\rho g\!}$ and
$a_{\!0\!}\!\!=\!\!1$. Eq.~(\ref{energydensitybulk}) yields the EL
Eqs.:
\begin{eqnarray}
M(a_{0},\bx) \bzeta_{0}  - (1-2\epsilon a_{0}^2) \bzeta_{0}^{''} +
\epsilon^2
\bzeta_{0}^{''''} = 0 \label{EL1} \\
M(a_1,\bx) \bzeta_1 - (1-2\epsilon a_1^2) \bzeta_1^{''} + \epsilon^2
\bzeta_1^{''''} = 0 \label{EL2} \\
\text{where:}  \ \ a_0^2 \bzeta_{0}(\bx)^2 + a_1^2 \bzeta_{1}(\bx)^2 = 1 \label{2modesinextent} \\
\text{and:} \ \ M(a_i,\bx) \equiv a_i^4+1+\bsigma(\bx)a_i^2/\epsilon
\ . \label{defineM}
\end{eqnarray}
With the constraint (\ref{2modesinextent}),
Eqs.~(\ref{EL1},\ref{EL2}) become a $4^{th}$ order nonlinear ODE for
the function $\bzeta_{0}(\bx)$ (alternatively, $\bzeta_{1}(\bx)$),
which must be solved under the two BC (\ref{BC2modes}). Obviously,
two BC do not suffice to solve a $4^{th}$ order ODE. One may show,
however \cite{preparation}, that in the regime
\begin{equation}
\epsilon <<1  \ ; \ a_{1} << 1/\sqrt{\epsilon} \  ,
\label{regiem2ODE}
\end{equation}
the {\emph{minimal energy}} profile that satisfy the
BC~(\ref{BC2modes}) is determined, to leading order in $\epsilon$,
by the $2^{nd}$ order ODE obtained from Eqs.~(\ref{EL1},\ref{EL2})
after neglecting the $4^{th}$ derivatives and the terms $2\epsilon
a_i^2\bzeta_i^{''}$. The BC~(\ref{BC2modes}) are thus sufficient for
finding the minimal energy profile of the
form~(\ref{2modessolution}). Physically, this means that force
balance in the regime (\ref{regiem2ODE}) is dominated by tensile
forces ($-\bzeta_i^{''})$ and by the restoring forces associated
with variation of the pattern from its preferred wrinkling period
($M(a_i,\bx)\bzeta_i$), whereas bending forces that result from
variation along $\hat{x}$ ($\epsilon^2 \bzeta_i^{''''},2\epsilon
a_i^2 \bzeta_i^{''}$) are negligible. Let us analyze now the
asymptotic behavior of $\bzeta_{0}(\!\bx\!),\bzeta_{1}(\!\bx\!)$. In
the limit $\bx \!\to \!\bX_0$ (more generally, for $\bx \! \ll \!1$)
we expect $\bsigma(\bx) \!\!\to\!\!
-(a_{0}^2\!\!+\!\!a_{0}^{-2})\epsilon \!\!=\!\! -2\epsilon$ and from
Eq.~(\ref{defineM}): $M(a_{0},\bX_0) = 0, M(a_{1},\bX_0) = a_{1}^4 +
1 - 2 a_{1}^2 >0$. In this limit, linear analysis of Eq.~(\ref{EL2})
implies:
\begin{equation}
\! \bzeta_{1} \!=\! \frac{Q_{1}}{a_{1}}
\sinh[\sqrt{\!M\!(a_1,\bX_0)} (\bx \!- \bX_0)] \ ; \ \bzeta_{0} \!
\approx \! \frac{1-\frac{1}{2}a_{1}^2\bzeta_{1}^2}{a_0} ,
\label{asymp1}
\end{equation}
where $Q_{1} \ (\!\propto\! e^{-\bL}\!)$ is a constant that is
determined by the solution of the nonlinear
Eqs.~(\ref{EL1}-\ref{2modesinextent}). The analogous analysis in the
limit $\bx\to \bX_1=0$ relies on the assumption $M(a_0,\bX_1) <0$,
that will be justified below. Linear analysis of Eq.~(\ref{EL1})
near $\bx \!=\!\bX_1$  then yields:
\begin{equation}
\!\bzeta_{0} \!\!=\!\! \frac{Z_{1}}{a_{0}} \sin[\!\sqrt{\!\!-\!\!
M(a_0,\bX_1)} (\bx\!\!-\!\bX_1 \!)] ;  \  \bzeta_{1} \approx
\frac{1-\frac{1}{2}a_{0}^2\bzeta_{0}^2}{a_{1}} , \label{asymp2}
\end{equation}
where $Z_{1}$ is another constant. Some algebraic manipulations
using
Eqs.~(\ref{BC2modes},\ref{2modesinextent},\ref{asymp2},\ref{EL1},\ref{defineM})
yield:
\begin{eqnarray}
M_{\!-\!}(a_{1},a_{0},Z_{1}) \equiv M(a_{0},\bX_1) =
\frac{a_{0}^4+1-a_0^2(a_{1}^2+\frac{1}{a_{1}^2})}{1-Z_{1}^2(\frac{a_{0}}{a_{1}})^2}
\label{defineMR} \\
\sigma(\bX_1) = \epsilon [-(a_1^2+\frac{1}{a_1^2}) + \frac{1}{a_1^2}
\! Z_{1}^2 M_{\!-\!}(a_{1},a_{0},Z_{1})] \label{sigmaR}
\end{eqnarray}
With Eq.~(\ref{defineMR}), the asymptotes
(\ref{asymp1},\ref{asymp2}), denoted hence as $\bzeta_i^L(\!\bx\!)$,
$\bzeta_i^R(\!\bx\!)$, respectively, are fully described by the two
unknown constants $Q_{1},Z_{1}$. Although they are the leading terms
in asymptotic expansions whose radii of convergence are unknown (and
may even vanish), we found an excellent agreement between the
function obtained from $\bzeta_i^L$, $\bzeta_i^R$ through standard
matching procedure and numerical solution of
Eqs.~(\ref{EL1}-\ref{defineM}) (in the regime (\ref{regiem2ODE}))
for all parameters $a_0,a_1$ that we tested. Such matching analysis
is based on equating $\bzeta_{0}^L(\!\bx\!)$ and
$\bzeta_{0}^R(\!\bx\!)$, and their first and second derivatives at
an unknown point $\bx^*\!\!<\!\!\bX_1$. The three matching
conditions translate into 3 algebraic equations for the unknowns
$\bx^*,Z_1,Q_1$, from which expressions for $Q_1$ and $Z_1$ as
functions of $a_0,a_1$, valid in the limit $\epsilon \!\!\ll\!\!1$,
are derived. Evaluation of the resulting formulas \cite{preparation}
proves that $0\!\!>\!\!Z_{1}(a_{0}\!\!=\!\!1,a_{1})\!\!>\!\!-\!1$
for all $a_{1}\!\!>\!\!1$, and thus (see Eq.~(\ref{defineMR})),
confirming the assumption $M(a_0,\bX_1) <0$. A characteristic shape
of the form (\ref{2modessolution}) with $a_1\!\!=\!\!2$ is plotted
in Fig.~2.
\begin{figure}
\vspace{-10mm}
\includegraphics[height=2.5in,clip=]{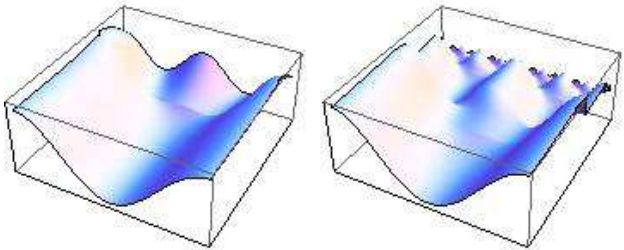}
\vspace{-15mm} \caption{\label{fig1} Computed patterns of wrinkled
membranes ($\epsilon \!\!\ll\!\!1$), with wavelength contrast
$\nu\!\!=\!\!1$ and a 1-strip shape (left), and with
$\nu\!\!=\!\!15$ and a 4-strip series $q_0(1,2,4,8,16)$ (right).}
\end{figure}
The energetic cost $U(\nu)$ (where $\nu \!\!=\! \!a_1\!-\!\!1$) of
the shape~(\ref{2modessolution}) with respect to the 1d periodic
wrinkling shape can be calculated (to leading order in $\epsilon$)
by substituting the obtained solutions in Eq.
(\ref{energydensitybulk}). We found the {\emph{linear}} behavior: $U
\!\!\approx\!\! (\nu\!-\!1)$, plotted as a gray solid line
(logarithmic scale) in Fig. 3.
\begin{figure}
\includegraphics[width=2.5in,clip=]{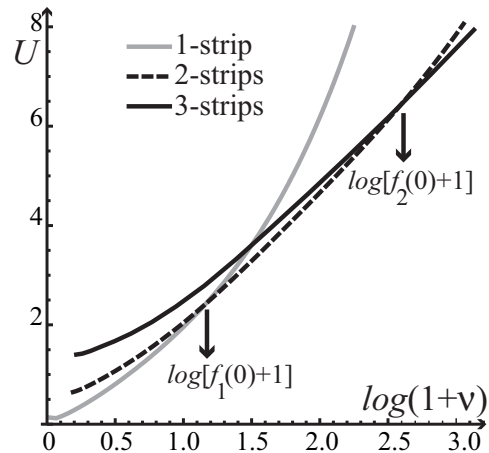}
\vspace{-3mm} \caption{\label{fig3} Energetic costs $U(\nu)$ (for
$\epsilon\!\! \ll \!\!1$) of symmetric (1-strip), 2-, and 3-strip
shapes (obtained by minimizing over all series
$(1\!\!>\!\!a_1\!\!>\!\!a_2\!\!>\!\!1\!+\!\nu)$. Dimensionless
convention, Eq.~(\ref{nondimensional}), is used.} \vspace{-5mm}
\end{figure}

The emergence of intermediate wavenumbers
$q_0\!\!<\!\!q_i\!\!<\!\!q_e$ in the wrinkling patterns observed in
\cite{Menon08} inspired us to study the stability of the symmetric
shape (\ref{2modessolution}) with respect to "$n$-strip" patterns. A
n-strip pattern is defined as $n$ consecutive strips parallel to
$\hat{y}$, with a series of wavenumbers
$(q_0\!<\!q_1\!\cdots\!<q_{n-1}\!<\!q_n=q_e)$, and a series of
borderlines $(X_{\!1\!},\cdots,X_{\!n\!-\!1\!})$ (with
$X_0\!\!=\!\!-\!L$ and $X_n\!\!=\!\!0$), such that the profile in
the $j^{th}$ strip is described by a superposition of the type
(\ref{2modessolution}) with the change of subscripts:
\begin{equation}
0\to j-1,1\to j \ . \label{substrans}
\end{equation}
As we now show, the shape in the $j^{th}$ strip, and in particular
its length $K_j \!\!=\!\! \bX_j \!\!-\!\! \bX_{\!j\!-\!1}$ are
determined iteratively from the shape in the $(\!j\!-\!1\!)^{th}$
strip by an energetic principle. Let us consider first the vicinity
of the borderline $\!\! \bX_{1} \!\!=\!\! -\!\!K_{\!1\!}$ in a
2-strip shape. We ask whether after the subscripts
change~(\ref{substrans}),the shape can be described by the
asymptotics~(\ref{asymp1}) (and $\bzeta_2(\bx)\!\!=\!\!0$) at $\bx
\!\!\to\!\! K_{\!1\!}^{\!-\!}$, and by the
asymptotics~(\ref{asymp2}) (and $\bzeta_0(\bx)\!\!=\!\!0$) at $\bx
\!\!\to\!\! K_{\!1\!}^{\!+\!}$, and a constant $Q_2$. Such a shape
implies a discontinuity of $\bzeta_{i}^{'}(\bx)$ at $\bx \!\!=\!\!
K_{\!1\!}$ and hence the divergence of $\bzeta_{i}^{''}(\bx)$, for
$i\!\!=\!\!0,2$. These divergences, however, stem from neglecting
the $4^{th}$ derivatives in Eq.~(\ref{EL1}), and are cured by the
formation of a boundary layer of size $\epsilon$ around
$\bx\!\!=\!\!K_{\!1\!}$ in which tensile and bending forces
($-\bzeta_i^{''},\epsilon^2 \bzeta_i^{''''}$, respectively) balance
each other. Similarly to the "take-off" line that is formed along a
paper sheet when it is pushed into a narrow ring, the energetic cost
of this layer is negligible for $\epsilon \! \!\ll \!\! 1 $
\cite{Witten06,preparation}. Let us consider now $\bzeta_{1}(\bx)$
near $K_{\!1\!}$. Eq.~(\ref{2modesinextent}) implies
$\bzeta_{1}^{'}(K_{1}) \!\!=\!\! 0$ and hence finite values of
$\bzeta_{1}^{''}(\bx)$ at both limits $\bx\!\!\to \!\!
K_{\!1\!}^{\!+\!},K_{\!1\!}^{\!-\!}$. If these values are different
the $4^{th}$ derivative $\epsilon^2\bzeta_{1}^{''''}(\bx)$ diverges,
leading to strong bending force that cannot be balanced by the
finite tensile force $\bzeta_{1}^{''}(\bx)$, and whose existence
thus leads to the emergence of a highly energetic region with
localized Guassian curvature, similarly to ridges on crumpled papers
\cite{Witten07}. The {\emph{``stitching condition"}}:
$\bzeta_{1}^{''}(\bx \!\!\to \!\! K_{\!1\!}^{\!-\!}) \!\!=\!\!
\bzeta_{1}^{''}(\bx\!\!\to\!\! K_{\!1\!}^{\!+\!})$, amounts to the
absence of such region. After manipulations similar to those that
led to Eq.~(\ref{defineMR}) this condition yields the asymptotics
(\ref{asymp1}) of the second strip at $\bx \!\!\to\!\!
K_{\!1\!}^{\!+\!}$ with:
\begin{eqnarray}
\! Q_{2} \!\equiv \! \sqrt{\frac{-(\frac{a_{0}}{a_2})^2
M_{\!-\!}(a_{1},a_{0},Z_{1}) Z_{1}^2 }{a_2^4 \!+\! 1 \!-\!
a_2^2(a_{1}^2 \!+ \! \frac{1}{a_{1}^2}) \!+ \!
(\frac{a_{0}}{a_{1}})^2 M_{\!-\!}(a_{1},a_{0},Z_{1}) Z_{1}^2}}
\label{defineQ} \\
M_{\!+\!}(a_2,a_1,Q_2)\equiv M(a_2,\bX_1) =
\frac{a_2^4+1-a_2^2(a_{1}^2+\frac{1}{a_{1}^2})}{1+Q_{2}^2(\frac{a_{2}}{a_1})^2}
\label{defineMplus}
\end{eqnarray}
The asymptotics of the second strip at $\bx
\!\!\to\!\!\bX_2\!\!\!=\! 0$ are given by
Eqs.~(\ref{asymp2},\ref{substrans}). The unknown constants are now
$K_{\!2\!}\!\!=\!\!\bX_2\!-\!\bX_1,Z_{\!2\!}$ which can be found by
following a matching procedure as described above, or by numerically
solving the $2^{nd}$ order ODE obtained from
Eqs.~(\ref{EL1}~-~\ref{2modesinextent},\ref{substrans}) in the
regime (\ref{regiem2ODE}) on an {\emph{unknown}} interval length
$K_{\!2\!}$ and {\emph{three}} BC:
$\bzeta_{1}(\bX_{\!2\!})\!=\!\bzeta_2(\bX_1)\!\!=\!\!0$ and
$\bzeta_2^{'}(\bX_{1}) \!\!=\!\! Q_{2}
\sqrt{M_{\!+\!}(a_2,a_1,Q_2)}$. One thus obtains functional
expressions for $K_{2},Z_{2}$ in terms of $a_{0},a_{1},a_2$. The
construction of $n$-strip with a series
$(a_{0}\!=\!1\!\!<\!\!a_1\!\!<\!\!\cdots\!\!<\!\!
a_{\!n\!-\!1\!}\!\!<\!\!a_n\!\!=\!\!1\!\!+\!\!\nu)$
proceeds iteratively: The shape at the $j^{th}$ strip is fully
determined from the known shape in the $(j-1)^{th}$ strip by an
identical procedure to the one described above where the subscript
change~(\ref{substrans}) is supplemented by $2\to j+1$. The energy
of the $n$-strip shape is then computed from
Eq.~(\ref{energydensitybulk}).

In order to analyze the stability of the symmetric (1-strip) shape
with respect to $n$-strip shapes, we calculated (for $\epsilon
\!\!\ll\!\! 1$) the {\emph{minimal}} energies of $2\!-\!\!$ and
$3\!-\!\!$~strip shapes for $\nu \!\!\in\!\!(0,20)$. Evaluation of
similar plots for $n\!\!>\!\!3$ requires elaborate computations,
since it involves (for every $\nu$) minimization in a high
dimensional ($d\!\geq\!3$) parameter space. We found that both
$1\!-\!\!$~strip and $2\!-\!\!$~strip energies scale
{\emph{logarithmically}} with $\nu$, such that $1\!-\!\!$~strip is
favorable for $0\!<\!\nu \!<\!f_1(0)$, $2\!-\!\!$~strip is favorable
for $f_1(0)\!<\!\nu \!<\!f_2(0)$, and $3\!-\!\!$~strip is favorable
(over $1\!-\!$ and $2\!-\!\!$~strip) for $\nu \!>\!f_2(0)$, where
$f_1(0) \!\!\approx \!\!3.2,\ f_2(0) \!\!\approx \!\!12$. These
observation leads to our second conjecture:

{\emph{{\bf{II.}} There exists a "period-fissioning" series
$\{\!f_n(\epsilon)\!\}$, such that for $(\!\epsilon,\nu\!)$ with
$\epsilon^{SI}(\!\nu\!) \!\! < \!\! \epsilon  \!\! < \!\!
(\!1\!\!+\!\!\nu\!)^2,f_n(\!\epsilon\!)\!\!<\!\!\nu\!\!<\!\!f_{n\!+\!1}(\epsilon\!)$,
the phase is $F_n$, characterized by $n$-strip shape.}}

In Fig.~4 we depict the conjectured irregular ($\{ I_n \}$),
symmetric ($S$) and period-fissioning ($\{F_n\}$) phases in the 2d
parameter space $(\epsilon_,\nu)$. As will be shown elsewhere
\cite{preparation}, the wrinkling patterns observed in
\cite{Menon08} seem to be described by phases $F_3$ and $F_4$.
\begin{figure}
\vspace{4mm}
\includegraphics[width=3in,clip=]{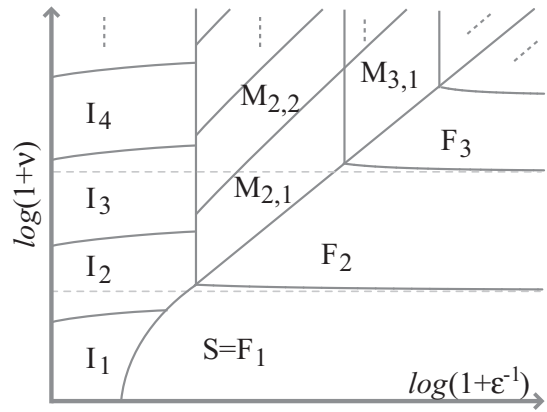}
\vspace{-2mm} \caption{\label{fig3} A conjectured phase diagram for
stressed membranes ($\tilde{\Delta}\!\!\ll\!\!1,\bL \!\!\gg\!\!1$).
The axis correspond to the dimensionless parameters
$(\epsilon,\nu)$. The irregular ($I_n$), period-fissioning ($F_n$),
and mixed ($M_{n,k}$) morphologies are described in the text.}
\vspace{-5mm}
\end{figure}
In order to understand the regime $\epsilon
\!\!>\!\!(\!1\!+\!\nu\!)^2$, consider
Eqs.~(\ref{sigmaR},\ref{substrans}). They predict that the
compression $|\sigma_{yy}(\bx)|$ in the $j^{th}$ strip increases
roughly as $q_j^2$, and hence imply that the {\emph{local}} value of
$\epsilon$ (compression-to-tension ratio) increases and becomes
$O(1)$ for $q_j \!\!\sim \!\!q_0/\sqrt{\epsilon}$. For wavelength
contrasts $\nu\!\!>\!\!\epsilon^{-1/2}$ one may thus expect
{\emph{mixed phases}} ($M_{n,k}$), characterized by $n$-strip that
terminates at $q_j \!\!\approx\!\! q_0/\sqrt{\epsilon}$, and
followed by an irregular shape $I_k$ that terminates at $q_e
\!\!=\!\! (1\!\!+\!\!\nu)q_0$.

The rich phase space depicted in Fig.~4 describes the morphology of
stressed membranes under rather restrictive conditions
($\tilde{\Delta} \!\! \ll \!\! 1$, constant tension, and BC mapped
onto a single number $\nu$). Future work will explore whether
morphologies beyond this regime are described by the conjectured
phases or whether new phases emerge. Finally, the structure depicted
in Fig.~4 raises the speculation that a nontrivial link exists
between the irregular ($\{I_n\}$) and period-fissioning ($\{F_n\}$)
phases. Computation of the conjectured series
$\{b_n(\epsilon\!),f_n(\epsilon\!)\}$ may reveal whether this
impression is merely a superficial one or points to a deep duality
between focusing and uniform distribution of stress in elastic
membranes. \vspace{-5mm}
\begin{acknowledgements}
\vspace{-3mm} The author is grateful to D.Nelson and B.Roman for
helpful comments, to H.Diamant, I.Dujovne and J.Machta for critical
reading of the manuscript, and to N.Menon and C.Santangelo for
valuable discussions.
\end{acknowledgements}


\begin{thebibliography}{99}
\bibitem{LLelasticity} L.D.Landau and L.M.Lifshitz, {\emph{Theory of
Elasticity}} (Pergamon,New York, 1986).
\bibitem{Cerda03} E.Cerda and L.Mahadevan, Phys. Rev. Lett. {\bf 90}, 074302
(2003).
\bibitem{Pomeau} Y.Pomeau, Phil. Mag. B {\bf 78}, 253 (1998); Y.Pomeau and S.Rica,
C.R. hebd. Seanc. Acad. Sci. Paris., Ser.IIb {\bf 325}, 181 (1997).
\bibitem{Menon08} J.Huang {\emph{et al.}}, submitted to Nature
Phyics.
\bibitem{Chicago08} L.Pocivavsek {\emph et al.}, Science {\bf 320}, 912
(2008).
\bibitem{Witten07} T.A.Witten, Rev. Mod. Phys. {\bf 79}, 643
(2007).
\bibitem{preparation} In preparation.
\bibitem{Witten06} T.Liang and T.A.Witten, Phys. Rev. E {\bf 71}, 016612
(2006).
\end{thebibliography}
\end{document}